\documentclass[sigplan,10pt]{acmart}
\startPage{1}

\setcopyright{none}
\usepackage{booktabs}   %% For formal tables:
\usepackage{subcaption} %% For complex figures with subfigures/subcaptions
\usepackage{syntax}

\usepackage{algorithm}% http://ctan.org/pkg/algorithms
\usepackage{algpseudocode}% http://ctan.org/pkg/algorithmicx

\begin{document}

%% Title information
\title[Short Title]{FusionStitching: Boosting Execution Efficiency of Memory Intensive Computations for DL Workloads}
                                        %% when present, will be used in
                                        %% header instead of Full Title.
%\titlenote{with title note}             %% \titlenote is optional;
                                        %% can be repeated if necessary;
                                        %% contents suppressed with 'anonymous'
%\subtitle{Subtitle}                     %% \subtitle is optional
%\subtitlenote{with subtitle note}       %% \subtitlenote is optional;
                                        %% can be repeated if necessary;
                                        %% contents suppressed with 'anonymous'

%% Author information
%% Contents and number of authors suppressed with 'anonymous'.
%% Each author should be introduced by \author, followed by
%% \authornote (optional), \orcid (optional), \affiliation, and
%% \email.
%% An author may have multiple affiliations and/or emails; repeat the
%% appropriate command.
%% Many elements are not rendered, but should be provided for metadata
%% extraction tools.

%% Author with single affiliation.
\author{Guoping Long, Jun Yang, Wei Lin}
%\authornote{with author1 note}          %% \authornote is optional;
%                                        %% can be repeated if necessary
%\orcid{nnnn-nnnn-nnnn-nnnn}             %% \orcid is optional
%\affiliation{
%  \position{Position1}
%  \department{Department1}              %% \department is recommended
%  \institution{Institution1}            %% \institution is required
%  \streetaddress{Street1 Address1}
%  \city{City1}
%  \state{State1}
%  \postcode{Post-Code1}
%  \country{Country1}                    %% \country is recommended
%}
\email{guopinglong.lgp, muzhuo.yj, weilin.lw@alibaba-inc.com}  %% \email is recommended

%% Author with two affiliations and emails.
%\author{First2 Last2}
%\authornote{with author2 note}          %% \authornote is optional;
%                                        %% can be repeated if necessary
%\orcid{nnnn-nnnn-nnnn-nnnn}             %% \orcid is optional
%\affiliation{
%  \position{Position2a}
%  \department{Department2a}             %% \department is recommended
%  \institution{Institution2a}           %% \institution is required
%  \streetaddress{Street2a Address2a}
%  \city{City2a}
%  \state{State2a}
%  \postcode{Post-Code2a}
%  \country{Country2a}                   %% \country is recommended
%}
%\email{first2.last2@inst2a.com}         %% \email is recommended
%\affiliation{
%  \position{Position2b}
%  \department{Department2b}             %% \department is recommended
%  \institution{Institution2b}           %% \institution is required
%  \streetaddress{Street3b Address2b}
%  \city{City2b}
%  \state{State2b}
%  \postcode{Post-Code2b}
%  \country{Country2b}                   %% \country is recommended
%}
%\email{first2.last2@inst2b.org}         %% \email is recommended

%% Abstract
%% Note: \begin{abstract}...\end{abstract} environment must come
%% before \maketitle command
\begin{abstract}
%Text of abstract \ldots.
Performance optimization is the art of continuous seeking a harmonious mapping
between the application domain and hardware. Recent years have witnessed a surge
of deep learning (DL) applications in industry. Conventional wisdom for 
optimizing such workloads
mainly focus on compute intensive ops (GEMM, Convolution, etc). Yet we 
show in this work, that the performance of memory intensive computations is vital
to E2E performance in practical DL models.

We propose \emph{FusionStitching}, a optimization framework capable of fusing 
memory intensive
\emph{elementwise}, \emph{reduction} and fine grained \emph{GEMM/Batched-GEMM} ops,
with or without data dependences, into large computation units, then mapping and 
transforming them into efficient GPU kernels. We formulate the fusion plan optimization
as an integer linear programming (ILP) problem, and propose a set of empirical
heuristics to reduce the combinatorial search space. In order to map optimized fusion
plans to hardware, we propose a technique to effectively compose various groups of
computations into a single GPU kernel, by fully leveraging on chip resources like
scratchpads or registers.
Experimental results on six benchmarks and four industry scale practical models are
encouraging. Overall, \emph{FusionStitching} can reach up to 5.7x speedup compared
to Tensorflow baseline, and achieves
1.25x to 1.85x performance speedups compared to current state of the art, with 1.4x on
average (geometric mean).
\end{abstract}

%% 2012 ACM Computing Classification System (CSS) concepts
%% Generate at 'http://dl.acm.org/ccs/ccs.cfm'.

%% End of generated code

%% Keywords
%% comma separated list
  %% \keywords are mandatory in final camera-ready submission

%% \maketitle
%% Note: \maketitle command must come after title commands, author
%% commands, abstract environment, Computing Classification System
%% environment and commands, and keywords command.
\maketitle

\section{Introduction}
%Text of paper \ldots
Recent years have witnessed a surge of industry scale applications of DL models,
ranging from text/NLP, audio/speech, images/videos, to billion scale search and
recommendation systems\cite{jizhe}. Such workloads are typically expressed with high 
level Python
APIs, modeled as computation DAGs, and mapped to hardware accelerators (such as GPUs)
through domain specific execution frameworks (such as Tensorflow\cite{tensorflow}, 
PyTorch\cite{torch}, Mxnet\cite{mxnet}, etc).
The challenge is how to transform high level computation graphs into efficient kernels
in order to maximize the execution efficiency on hardware.

In DL workloads, dense tensor computations (GEMMs, Convolutions, etc) are ubiquitous.
Thus many prior research works either focus on optimizing performance of such compute 
intensive primitives\cite{halide,tc,tvm}, or target a kernel selection and computation
scheduling problem\cite{astra,pbqp}. This approach works well for workloads that are
dominated by FLOPs efficiency of GEMMs or Convs, for instance, convolutional neural
networks\cite{vgg,resnet,inception}.
However, recent advancement of the DL domain has resulted in many novel model structures
which are dominated by memory intensive patterns (element wise layers, layout transpose
or reduction
operators). In addition, there are models with a large number of fine grained (< 10us) 
operators,
causing notable runtime launch overheads when executing on GPUs. For these workloads,
optimizing compute
intensive ops alone is inadequate to unlock the full potential of execution efficiency.

To this problem, libraries such as cuDNN/cuBlas have the capability to fuse element 
wise layers into large compute intensive kernels. It is also possible to fuse multiple
GEMMs into a large GEMM or a batched GEMM op\cite{astra}. However, these techniques do 
not work well for workloads that are dominated by memory intensive structures.

Another known approach is kernel fusion, a technique to fuse multiple memory 
intensive ops with data dependencies into a single kernel to reduce off chip memory
accesses.
Prior works have explored this idea extensively in database\cite{kernelweaver}, 
image processing\cite{cgo2019, halide, video}, HPC applications\cite{hpcfusion, ode-fusion}, 
and AI workloads\cite{xla,ppopp2015}. However, there are two notable limitations
when targeting memory intensive DL models.
First, current techniques mainly focus on reducing memory traffic instead of reducing
the number of kernel launches, therefore give up fusing as many computations
with no data dependences as possible. Second, existing works only compose
elementwise or reduction layers, lacking the ability to fuse and optimize
compute and memory intensive ops together.

In this work, we propose \emph{FusionStitching},
an optimization framework to systematically perform fusion space
exploration and aggressive code generation. One key observation underpinning our
methodology is the inherently repetitive nature of DL workloads. That is, by optimizing
thoroughly once, it is possible to reuse the optimized implementation in future runs.

\emph{FusionStitching} addresses limitations faced by current state of the art by
composing as many fine grained ops as possible, with or without data dependences,
into large GPU kernels, in order to reduce off chip memory accesses and launch 
overheads simultaneously. In particular, it comes with the capability
to optimize compute and memory intensive ops in a uniform optimization scope,
and perform global fusion and code generation in entirety. Besides, we make three
unique contributions to the community:
\begin{itemize}
\item[-] Naive composition of multiple computations may cause notable performance
slowdown, because different portions of the kernel may have conflicting memory
layout, parallelization or on chip resource requirements\cite{versapipe}. 
In this work, we formulate the fusion 
plan optimization as an integer programming problem. In addition, we introduce
effective domain specific heuristics to make the solution space exploration 
practically tractable.
\item[-] To the best of our knowledge, this is the first work to integrate compute
intensive ops into the fusion and kernel generation of memory intensive computations.
\item[-] We propose a lightweight, domain specific representation of potentially
complex kernel implementations. The compact representation enables separation
between the kernel optimization intent and implementations, and facilitates parametric
performance tuning effectively.
\end{itemize}

Experimental results on six benchmarks and four practical industry models show 
promising results in both kernel compression ratio (kernel number reduction) and
performance speedup. Compared to the Tensorflow XLA\cite{xla} baseline, 
\emph{FusionStitching}
can achieve up to 10x kernel compression ratio, with 2.8x on average.
As to performance, \emph{FusionStitching} can reach up to 5.7x speedup compared
to Tensorflow baseline, and achieves 1.25x to 1.85x performance speedups 
compared to XLA, with 1.4x on average.
Please note that although all benchmarks are expressed with Tensorflow APIs, 
our approach applies to other DL frameworks (such as MxNet, PyTorch, etc) as well. 

The rest of the paper is organized as follows. Section \ref{section:motivation}
presents characterization of DL workloads that motivate this work. Section
\ref{section:overview} presents the high level sketch of our approach. Section
\ref{section:fusion} and \ref{section:codegen} presents our fusion
and kernel generation mechanisms, respectively. Section \ref{section:evaluation}
discusses experimental results. Section \ref{section:relatedwork} discusses related
works and section \ref{section:conclusion} concludes this work.
\section{Motivation}
\label{section:motivation}
%Text of paper \ldots
We present key observations of emerging DL workloads to highlight the significance 
of memory intensive ops. Table \ref{tbl:benchmark-summary} summarizes our workloads,
including 4 production models under deployment at Alibaba, and 6
micro-benchmarks from Tensorflow-Examples\cite{tf-examples}. Application domains of
these benchmarks are diversified, ranging from basic ML algorithms
(\emph{logistic}, \emph{word2vec}, \emph{rnn}, \emph{perceptron}, \emph{var-encoder}), 
NLP (\emph{nmt}\cite{NMT}, \emph{aiwriter}), audio (\emph{rokid}),
to internet scale E-commerce search and recommendation systems (\emph{multi-interests}
\cite{recommender,jizhe}).
Among them, \emph{nmt} is an inference application, and all others are training models.
\begin{table}[t]
\caption{Workload Description}
\label{tbl:benchmark-summary}
\begin{center}
\begin{small}
%\begin{sc}
\begin{tabular}{lll}
\toprule
Category & Name & Description\\
\midrule
Application & nmt & Neural machine translation\\
Application & multi-interests & Recommender systems\\
Application & rokid & Speech recognition\\
Application & aiwriter & Dialog generation\\
Micro-Benchmark & logistic & Logistic regression\\
Micro-Benchmark & word2vec & Word to vector embedding\\
Micro-Benchmark & bi-rnn & Bi-directional RNN\\
Micro-Benchmark & dyn-rnn & Dynamic RNN\\
Micro-Benchmark & perceptron & Multilayer neural network\\
Micro-Benchmark & var-encoder & Variational encoder\\
\bottomrule
\end{tabular}
%\end{sc}
\end{small}
\end{center}
\vskip -0.1in
\end{table}

Table \ref{tbl:benchmark-characteristics} presents key characteristics of our workloads.
\emph{\#Graph Size} denotes the number of Tensorflow ops of the running target graph.
\emph{\#Kernels} denotes the number of GPU computation kernels. In this context, we do 
not consider data transfer kernels back and forth between CPU and GPU, because all 
workloads are dominated by computation. \emph{Avg. Kernel Size} shows average kernel
execution time, a metric to measure kernel granularity. \emph{Mem. Ratio} denotes
the percentage (exe. time) of memory intensive kernels in all computation kernels.

As can be seen, in all workloads except \emph{multi-interests}, average kernel time is
less than 10us. This is close or even smaller than a single driver 
launch latency\cite{astra,nvdoc}.
Even for \emph{multi-interests}, close examination reveals 
interesting possibilities to optimize compute and memory intensive computations from
global perspective (Section \ref{section:overview}). 
Besides fine grained kernels, there are up to 60\% on average of \emph{Mem. Ratio} 
for all workloads. These observations motivate \emph{FusionStitching}, our solution
to boost the execution efficiency for memory intensive computations.
\begin{table}[t]
\caption{Workload Characteristics}
\label{tbl:benchmark-characteristics}
\begin{center}
\begin{small}
%\begin{sc}
\begin{tabular}{lllll}
\toprule
Name & \#Graph Size & \#Kernels & Avg. Kernel Size & Mem. Ratio\\
\midrule
nmt & 1532 & 5940 & 9.2us & 46\% \\
multi-interests & 992 & 495 & 202.8us & 81\% \\
rokid & 3204 & 261029 & 4us & 47\% \\
aiwriter & 27428 & 398265 & 5.4us & 80\% \\
logistic & 62 & 1540 & 3.1us & 60\% \\
word2vec & 206 & 64 & 2.5us & 71\% \\
bi-rnn & 461 & 1085 & 3.7us & 56\% \\
dyn-rnn & 957 & 1352 & 3.2us & 63\% \\
perceptron & 108 & 49 & 5.4us & 47\% \\
var-encoder & 239 & 112 & 4.5us & 53\% \\
\bottomrule
\end{tabular}
%\end{sc}
\end{small}
\end{center}
\vskip -0.1in
\end{table}

\section{Overview}
\label{section:overview}
\subsection{A Motivating Example}
\begin{figure}
\includegraphics[scale=.2]{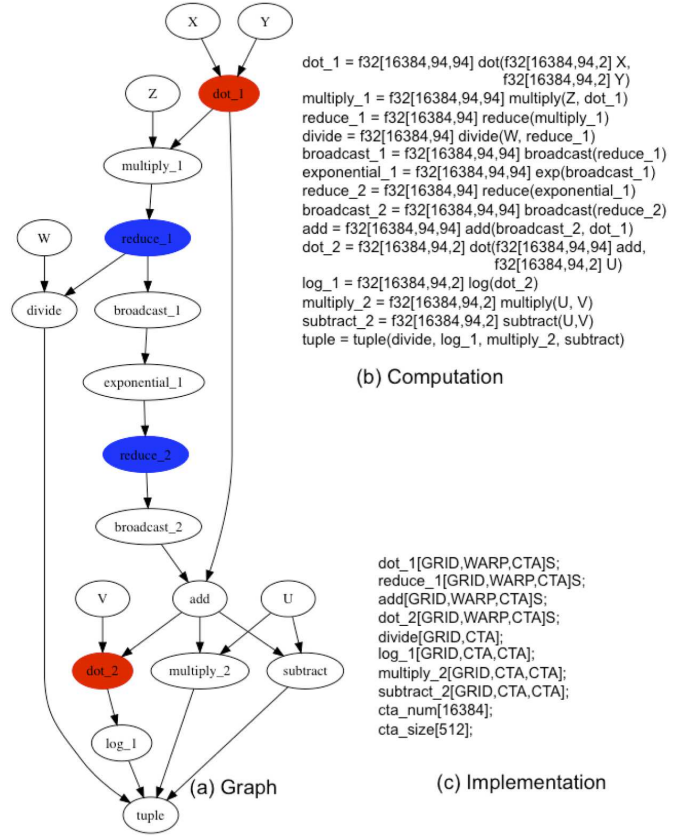}
\caption{A Motivating Example}
\label{fig:example}
\end{figure}

%%Text of paper \ldots
In this section, we present a motivating example from \emph{multi-interests}. It is
a pattern generated by the fusion process (Section \ref{section:fusion}), with
irrelevant details omitted for illustration purposes. Figure \ref{fig:example} (a)
and (b) show the data flow graph and computation specification, respectively. 
Here we adopt the XLA tensor operation 
semantics\cite{semantics} in our presentation.

This example illustrates the necessity to optimize compute and memory intensive
ops collectively. It includes a combination of nine elementwise, two reduction and 
two batched
GEMM (\emph{dot\_1} and \emph{dot\_2}) ops. The terminal \emph{tuple} groups all
output tensors (\emph{divide}, \emph{log\_1}, \emph{multiply\_2} and \emph{subtract})
of this fused computation. In particular, \emph{dot\_1} takes two
small tensors (12MB) as input and generate a large (552MB) output tensor. Finally,
\emph{dot\_2} takes the large tensor originated from \emph{dot\_1} and produces a 
small output. Fusing all these ops into a kernel reduces off chip memory accesses
substantially caused by intermediate results.

We highlight three design principles to high performance kernel generation for
such complex fused computations.
First, the work space of the entire computation must be parallelizable across shape
dimensions. Second, on chip memory is preferable to transfer intermediate 
results, thus avoid redundant computation. Third, fast memory (such as registers)
is preferred over relatively slow storage (such as shared/L2 memory) for computation
reuse.
Figure \ref{fig:example} (c) shows an implementation template. It defines one
possible implementation of the fused kernel. This is achieved by specifying how 
to parallelize the work space (through \emph{GRID,CTA,WARP}), how to transfer
intermediate results (through \emph{S} attribute), and kernel launch parameters
(through \emph{cta\_num} and \emph{cta\_size}). We present details of kernel
generation in Section \ref{section:codegen}.

\subsection{Overview}
\begin{figure}
\includegraphics[scale=.4, bb=60 87 622 387]{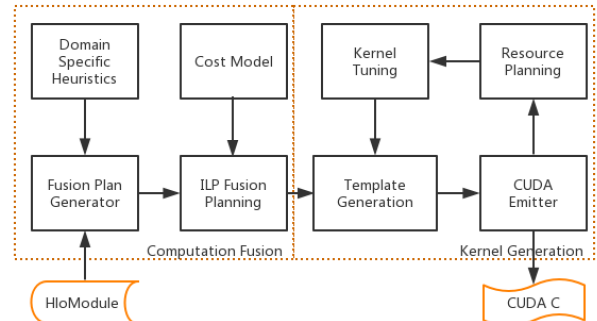}
\caption{Overview}
\label{fig:overview}
\end{figure}

Figure \ref{fig:overview} shows the overview of our approach. In the fusion phase,
the fusion plan generator takes domain specific heuristics, and generate a modest
number of fusion patterns for evaluation. The fusion planner takes a cost
model and produce a score for each pattern. Then the ILP solver outputs an
optimized fusion plan for the entire computation, which is executed to generate
a fused representation for kernel generation.

In the kernel generation phase, a template generator produces many templates, each
with different trade-offs among parallelization, intermediate results sharing and
launch settings. Since a template uniquely identifies a kernel, the kernel tuner
can iterate over the implementation space by traversing these compact templates.
Given a template, the resource planner orchestrates on chip resource usage. Instead
of generating LLVM IR representation directly, the CUDA emitter produces a CUDA 
C kernel for diagnosis. With increasingly complex fusion patterns, this greatly 
ease debuggability and performance tuning as well.

\section{Computation Fusion}
\label{section:fusion}
%Text of paper \ldots
\subsection{The ILP Formulation}
Given a computation graph $G=(V,E)$, with $V$ and $E$ are sets of vertices and 
edges, respectively. We define a fusion pattern $P_i=(V_i,E_i)$ as a subgraph of
$G$, with $V_i \subseteq V$, $E_i \subseteq E$. For each fusion pattern $P_i$, we 
define an integer variable $X_i$ where $0 \le X_i \le 1$, and a real score
function $f(X_i)$ where $f(X_i) \ge 0$. The larger the score $f(X_i)$, 
the higher the fusion gain w.r.t. performance. 
Given a set of fusion patterns $S=\{P_1,\ldots,P_k\}$, we have to
\begin{equation*}
\begin{aligned}
\text{maximize} \sum_{j=1}^{k}X_j f(P_j) \\
\text{subject to:}  X_u+X_v \le 1, \forall u,v, 0 \le u,v \le k, P_u \cap P_v \ne \emptyset.
%\text{subject to} \sum_{j=1}^{n}w_j x_j \leq c, x_j \in \{0,1\}, j = 1, \ldots, n.
\end{aligned}
\end{equation*}
We define a fusion plan of a computation graph as a subset of $S$.
Informally, the objective is to resolve a fusion plan $S^-$ in order to maximize
the total fusion gain, such that any two fusion patterns in $S^-$ are disjoint. That is,
for any node in the graph, it can only exist in at most one fusion pattern.

We require that the score $f(X_i) \ge 0$. This means we give up all fusion patterns 
with negative gains. In theory, multiple patterns with negative scores may be fused 
together to produce a composite one with positive gain. There are two things to note. 
First, for practice, our design rationale is to cover known performance critical patterns 
in DL workloads, rather than exhausting all combinatorial possibilities, a prohibitively
expensive operation. Second, even if we do not perform exhaustive search, there is still
high probability that our domain specific pattern generator can discover the composite
pattern. We discuss domain specific fusion space explorations in the next subsection.

\begin{figure}
\includegraphics[scale=.3]{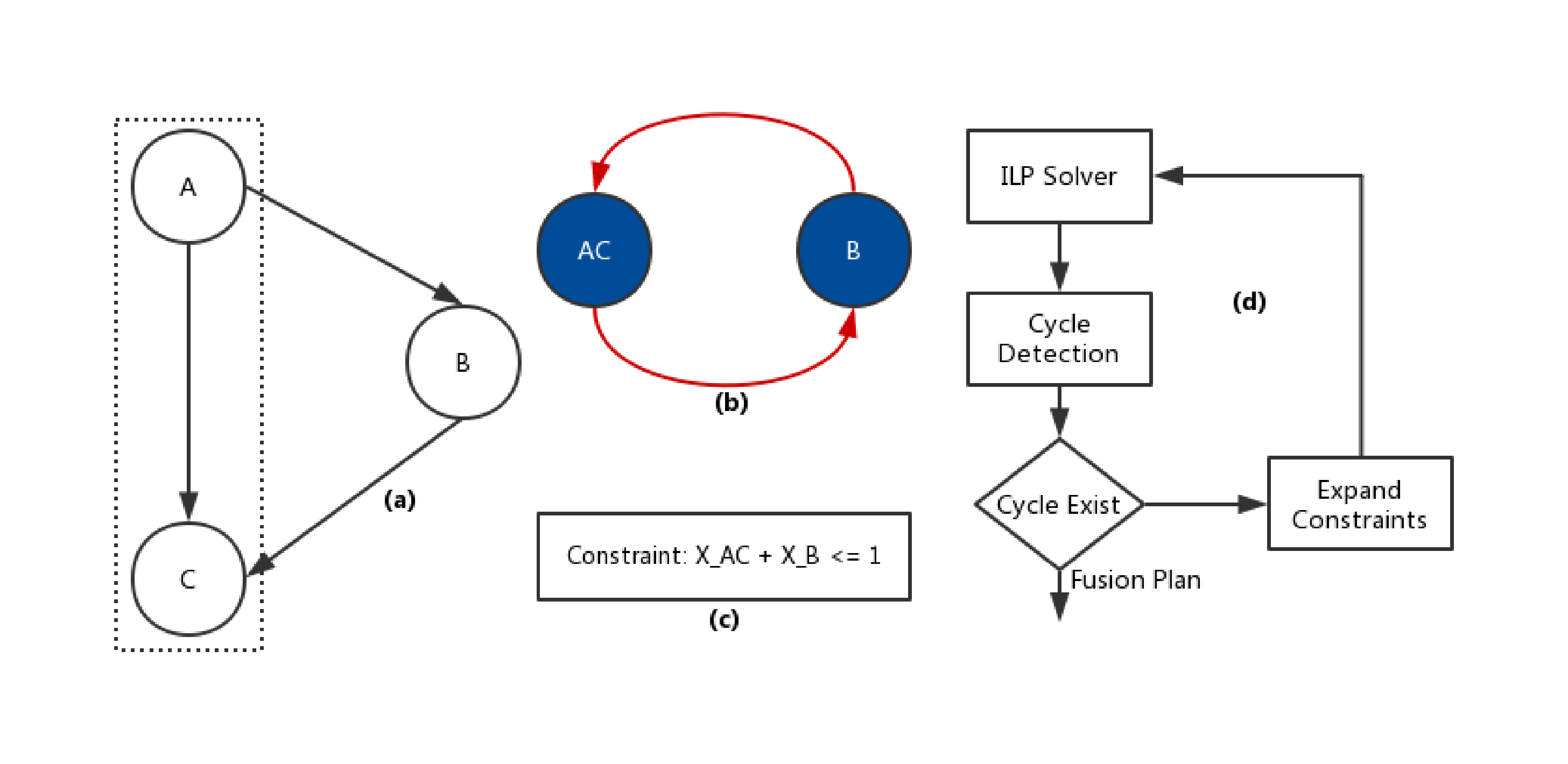}
\caption{Fusion Plan: Cyclic Dependence}
\label{fig:cycle}
\end{figure}

Note that the computation graph is a DAG. We must ensure that, after executing the 
fusion plan,
the resultant graph is also a DAG. However, the ILP solver alone does not enforce
this property, as shown in Figure \ref{fig:cycle}. In this simple graph 
(Figure \ref{fig:cycle}(a)), nodes $A$ and $C$ are fused together, thus causing a
cyclic dependence (Figure \ref{fig:cycle}(b)).

We propose an iterative process to address this issue (Figure \ref{fig:cycle}(d)).
When the ILP solver generates a fusion plan, we check if there exists a cycle. If so,
we identity the cycle, and enrich the solver with additional constraints 
(Figure \ref{fig:cycle}(c)). We repeat the process until there is no cycle. The final
fusion plan is used to transform the computation graph before kernel generation.

\subsection{Domain Specific Pattern Generation}
In this work, we consider fusion patterns consisting of \emph{gemm},
\emph{batched-gemm}, \emph{reduction} and \emph{elementwise} ops, 
with or without data dependences among them.
Given a computation graph, a naive approach to generate fusion patterns is to
iterate over the entire permutation space by evaluating each subgraph. 
This is too slow to operate in practice.
We present two pattern search algorithms from DL domain specific
perspective. The goal is to capture important fusion opportunities by traversing 
a modest subset of the search space.
\subsubsection{Substitution Fusion}
The substitution fusion targets graphs with tiny kernels, as shown in most models
in Table \ref{tbl:benchmark-summary}. Algorithm \ref{alg:substitute-fusion} shows
the procedure: it takes the $graph$ and a $partition\_ops$ as input, and outputs 
fusion patterns. The $partition\_ops$ identifies ops that will
never be fused with other ops. The basic idea is to minimize the number of kernels.
To achieve this, we first topologically sort the graph and $partition\_ops$. 
Then we iterate over the sorted list ($sorted\_parts$), and collapse all ops between
adjacent partition ops into a single fusion pattern.

For DL workloads, we use a multi-step heuristic to identify $partition\_ops$ and
produce the full set of $fusion\_patterns$ for the ILP solver. First, we add all
large \emph{gemm} ops into $partition\_ops$, and run 
Algorithm \ref{alg:substitute-fusion} to generate fusion patterns. Next, we 
extend $partition\_ops$ to further include \emph{batched-gemm}, and again
run the algorithm to obtain more fusion patterns. We then repeat this step to
further include \emph{column reductions}, \emph{scalar reductions}, etc. Each
time we extend the $partition\_ops$, we run the algorithm to generate more fusion
patterns, until $partition\_ops$ can not be extended anymore.

The motivation of this multi-step procedure stems from our observations in 
optimizing DL models. Given code generation capability, fusing multiple ops may
or may not produce performance gains. Except \emph{elementwise} or 
\emph{row reduction} patterns, all others
have dedicated requirements on parallelization strategies, on chip resource 
requirements, and launch constraints. These requirements may conflict with each
other when aggressively composing them together.

The substitution fusion typically produces a minimum number of kernels. It also
enjoys the property of being cycle free, because we never fuse across the 
partition op boundary. However, this may miss some interesting optimizations, 
such as the case shown in Figure \ref{fig:example}. The exploratory fusion fixes 
this problem.
\begin{algorithm}[tb]
   \caption{The Substitution Fusion Algorithm}
   \label{alg:substitute-fusion}
\begin{algorithmic}
   \State {\bfseries Input:} $graph$, $partition\_ops$
   \State {\bfseries Output:} $fusion\_patterns$
   \State $topo\_order=TopologicalSort(graph)$
   \State $sorted\_parts=TopoSortPartOps(topo\_order, partition\_ops)$
   \For{$p$ {\bfseries in} $sorted\_parts$}
   \State $pattern=AllOpsUpToNextPartOp(p,topo\_order)$
   \State $fusion\_patterns.append(pattern)$
   \EndFor
\end{algorithmic}
\end{algorithm}

\subsubsection{Exploratory Fusion}
Unlike the substitution fusion, the exploratory fusion targets \emph{elementwise},
\emph{batched-gemm} or \emph{reduction} patterns with large granularity. The 
objective is to minimize off chip memory accesses by composing as many data 
dependent ops as possible. Algorithm \ref{alg:explore-fusion} shows the procedure.
It is a recursive process taking the computation $graph$ and $seed\_pattern$, 
an initial set of ops for fusion, as input, and producing fusion patterns for 
the ILP solver. The producer (consumer) expansion examines all producers (consumers)
of $seed\_pattern$ and put all fusible ops into $candidates$. We consider an $op$ that 
can be fused into $seed\_pattern$ only if two conditions are both satisfied: 
(1) $op$ must be a \emph{elementwise}, \emph{batched-gemm}, \emph{reduction} op; 
(2) fusion of $op$ into $seed\_pattern$ does not introduce any cyclic data dependence.
After candidates expansion, we examine each op $p$ in $candidates$ by
generating a new fusion pattern $fusion$ by fusing $p$ into $seed\_pattern$. Then we 
explore starting from the expanded $fusion$ recursively. 

Naive application of the exploratory fusion risks exploring a huge search space.
We mitigate this issue with careful selection of the initial $seed\_pattern$.
Specifically, we consider two heuristics. First, ops with large ($>10$, for instance)
number of operands are excluded, as Algorithm \ref{alg:explore-fusion} explores 
all subsets of operands. Second, we only consider \emph{elementwise}, 
\emph{batched-gemm}, \emph{reduction} ops with large input/output tensors.

In practice, we use the substitution fusion as a base strategy, and the exploratory
algorithm as supplementary. If after applying these heuristics, the exploration still
takes long time, we simply give up further exploration. While this may miss some
marginal fusion possibilities. They are not significant to performance in all our
workloads.
\begin{algorithm}[tb]
   \caption{The Exploratory Fusion Algorithm}
   \label{alg:explore-fusion}
\begin{algorithmic}
   \State {\bfseries Input:} $graph$, $seed\_pattern$
   \State {\bfseries Output:} $fusion\_patterns$
   \Procedure{Explore}{$graph$,$seed\_pattern$}
     \State {\bfseries Initialize empty set:} $candidates$
     \State $ProducerExpansion(graph,seed\_pattern,candidates)$
     \State $ConsumerExpansion(graph,seed\_pattern,candidates)$
     \For{$p$ {\bfseries in} $candidates$}
       \State $fusion=Fuse(seed\_pattern,p)$
       \State $fusion\_patterns.append(fusion)$
       \State {\Call{Explore}{$graph$,$fusion$}}
     \EndFor
   \EndProcedure
\end{algorithmic}
\end{algorithm}

\subsection{Fusion Pattern Evaluation}
Fusion pattern evaluation measures the performance gain of the fused computation.
It produces a real valued score for each pattern.
A negative score means either we can not generate a kernel for the pattern, 
or the fused kernel performance is less than satisfactory. Thus only patterns 
with non-negative scores are fed into the ILP solver.

There are two techniques to evaluate fusion patterns. One is execution based.
That is, we run the kernel generator to produce a kernel for the fused pattern,
and measure its performance directly. Given a fusion pattern $P=\{Op_1,\ldots,Op_N\}$,
the score is defined as follows:
\begin{equation*}
\begin{aligned}
f(P)=\sum_{j=1}^{N}K(Op_j) + (N-1)*\phi - K(P)
\end{aligned}
\end{equation*}
Here $K(Op_j)$ and $K(P)$ denote kernel execution time of $Op_j$ and the fusion 
pattern $P$, respectively. $\phi$ is the average launch latency (between $6us$ and 
$10us$) for a kernel. $f(P)$ measures the execution time saved after fusion. The
above formula assumes the fused kernel has been generated successfully. If not so,
we simply return a negative $f(P)$ value, and the fusion pattern is ignored.

\begin{figure}
\includegraphics[scale=.4]{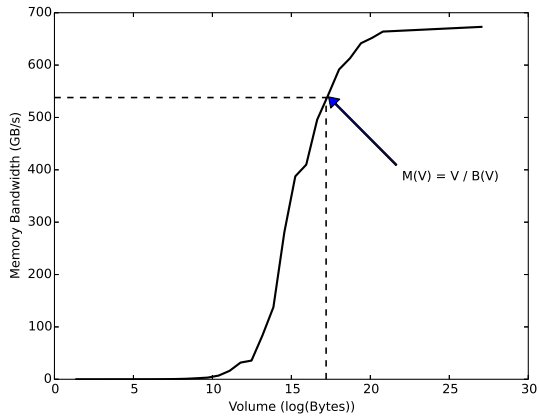}
\caption{Memory Bandwidth Model}
\label{fig:model}
\end{figure}

The other approach is model based. Since we allow general composition of
\emph{gemms}, \emph{batched-gemms}, \emph{reductions} and \emph{elementwise} ops,
on chip shared memory is essential to transfer intermediate results (Section 
\ref{section:codegen}). We first check if the shared memory usage is satisfiable.
If not, we simply return a negative score and ignore the pattern. Otherwise,
we measure $V$, the volume (in Bytes) of input/output off chip memory accesses 
saved. Given fusion pattern $P=\{Op_1,\ldots,Op_N\}$, the score is thus defined 
as follows:
\begin{equation*}
\begin{aligned}
f(P)=M(V) + (N-1)*\phi
\end{aligned}
\end{equation*}
Here, $M(V)$ models the extrapolated latency accessing $V$ bytes of memory. We make
two simplifications to enable fast calculation of $M(V)$. First, we assume consecutive
access of $V$ bytes of memory. Second, $M(V)$ depends on hardware memory bandwidth
capability and the size of $V$. In order to avoid measuring $M(V)$ for each $V$,
we use an memory bandwidth utilization model collected offline on the same hardware.
We thus extrapolate $M(V)$ using the model, as shown in Figure \ref{fig:model}.

The model based evaluation is fast, yet with less accuracy for complex fusion patterns.
On the contrary, the execution based approach is accurate by generating, executing,
and profiling the target kernel directly, but is time consuming. We adopt a combination
of both. Specifically, we use model based approaches for most memory intensive patterns,
but enable the execution based approach for complex ones. By complex, we mean domain
specific heuristics, for instance fusion patterns with combinations of column/scalar and
row reductions, gemm/batched-gemms and reductions, etc.
\subsection{Implementation}
We implement the fusion mechanism together with cost models as a code transformation 
pass in the XLA compilation framework in Tensorflow. As to the ILP solver, we use the
publicly available Python \emph{pulp} package.
\section{Kernel Generation}
\label{section:codegen}
%Text of paper \ldots
\subsection{Computation Composition}
\begin{figure*}
\centering
\includegraphics[scale=.42, bb=47 81 1268 579]{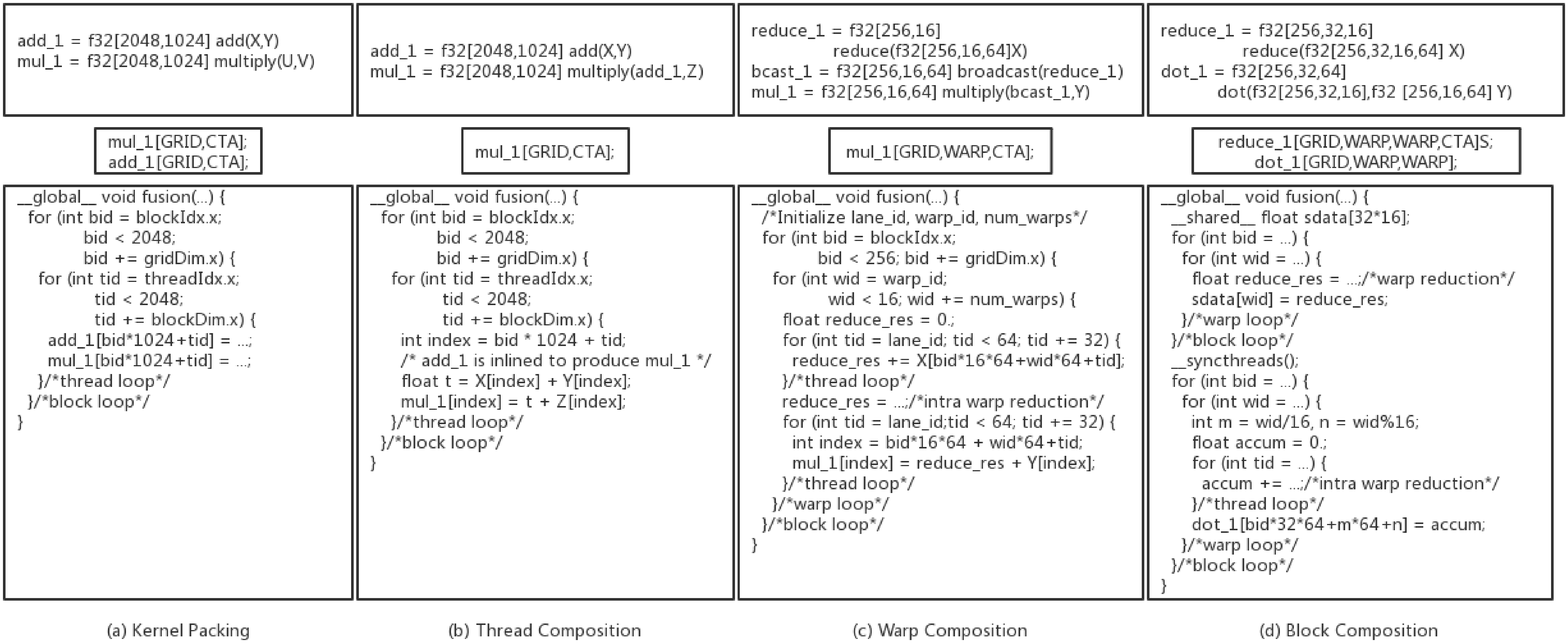}
\caption{Computation Composition}
\label{fig:composition}
\end{figure*}
Given a fusion pattern, a key issue in kernel generation is how to compose computations
of multiple ops into a fused kernel. Figure \ref{fig:composition} summarizes four types
of composition mechanisms supported in our work. For each composition, we show a 
computation example, the implementation template (Section \ref{section:template}), 
and the generated kernel sketch.

Kernel packing (Figure \ref{fig:composition} (a)) packs computations of ops with 
no data dependences. Two observations of DL workloads motivate this. One is the
existence of very fine grained ops as shown in most workloads in Table
\ref{tbl:benchmark-characteristics}. Another is related to the backward phase
of DL training. Groups of independent elementwise layers with similar or even
identical shapes operate on large tensors for gradients accumulation.
The kernel sketch includes two nested parallel loops, one for thread
blocks and the other for threads within a block. To reduce control flow overheads,
we perform aggressive loop fusion \cite{Kennedy,gao} to merge as many elementwise ops
as possible into a single loop structure. While the example only illustrates
packing of elementwise ops, we also pack reduction and compute intensive ops
as well. For DL workloads, kernel packing is instrumental in reducing loop control
and kernel launch overheads.

Thread composition (Figure \ref{fig:composition} (b)) fuses data dependent ops
within a local thread context. Intermediate results are transferred via registers.
For modern GPUs with large register files, this enables composition of many
elementwise ops into large fused kernels.
Warp composition (Figure \ref{fig:composition} (c)) extends thread composition
to fuse elementwise ops with a special form of reduction patterns. Such structures
exist in common DL building blocks such as \emph{softmax}, \emph{batchnorm}, 
\emph{layernorm} structures and their variants. Here we fuse the block loop,
and employ warp reduction to enable register transfer of intermediate results of 
\emph{reduce\_1} to dependant elementwise ops (\emph{mul\_1}).

Block composition (Figure \ref{fig:composition} (d)) is essential to compose
\emph{elementwise}, \emph{reduction} and \emph{gemm/batched-gemm} ops into a 
fused kernel. Unlike registers, we use on chip shared (\emph{scratchpad} memory 
to transfer intermediate results. This is flexible, because we allow different ops
(\emph{reduce\_1} and \emph{dot\_1}) to have independent parallel loops. Block
composition unlocks the potential to enable composing non-homogenous computations
into large fused kernels.

Previous works explored thread and block compositions in database\cite{kernelweaver},
image processing\cite{cgo2019, halide, video}, and HPC applications\cite{hpcfusion, ode-fusion}.
The Tensorflow XLA\cite{xla} framework implements thread composition. 
To the best of our knowledge, this work is the first to investigate
all four computation compositions thoroughly for DL workloads.
\subsection{The Implementation Template}
\label{section:template}
We propose the implementation template as a compact representation of concrete cuda 
kernels. Given a fusion pattern and such a template, the code generator synthesizes 
a kernel (Section \ref{section:kernelgen}). 
Following is the grammer for templates specification:
\setlength{\grammarparsep}{2pt plus 1pt minus 1pt} % increase separation between rules
\setlength{\grammarindent}{9em} % increase separation between LHS/RHS 
\begin{grammar}
%<statement> ::= <ident> `=' <expr> 
%\alt `for' <ident> `=' <expr> `to' <expr> `do' <statement> 
%\alt `{' <stat-list> `}' 
%\alt <empty> 
%
%<stat-list> ::= <statement> `;' <stat-list> | <statement> 
<template> ::= <schedule-list>

<schedule-list> ::= <schedule> <schedule-list> | <schedule>
%\alt <schedule>

<schedule> ::= <identifier> `[' <attr-list> `]' `;'
\alt <identifier> `[' <attr-list> `]' `S' `;'

<attr-list> ::= <attr> `,' <attr-list> | <attr>
%\alt <attr>

<attr> ::= <attrtype> | <subattr-list>
%\alt <subattr-list>

<subattr-list> ::= <subattr> `-' <subattr-list> | <subattr>
%\alt <subattr>

<subattr> ::= <attrtype> `\_' <integer>

<attrtype> ::= `GRID' | `WARP' | `CTA' | `THREAD'
\end{grammar}

The template consists of one or more schedules. A schedule denotes an \emph{op}
implementation that writes either shared (intermediate results used by other ops)
or off-chip global memory (outputs of the entire fusion pattern). As an example,
let's consider \emph{reduce\_1} in Figure \ref{fig:composition} (d). In the schedule
\emph{[GRID,WARP,WARP,CTA]S}, there is a parallel loop tiling attribute for each input 
dimension. Here we perform block level parallelization on the first, warp level
parallelization on the second and the third, and thread level parallelization
on the last dimension. We also support multiple levels of tiling on the same 
dimension. For instance, in schedule \emph{[GRID\_128-WARP\_2,WARP,WARP,CTA]S},
we perform both block and warp level tiling on the first dim. The \emph{S} attribute
instructs the kernel generator to cache results in shared memory.

%Unlike \emph{GRID}, \emph{WARP} or \emph{CTA}, 
The \emph{THREAD} attribute has no parallelization
implications. For example, in schedule \emph{[GRID,CTA,CTA,THREAD]}, each thread
performs sequential reduction independently. This is useful when the reduction dim 
size is trivially small (not uncommon in practical workloads), applying this 
schedule essentially treats the reduction as elementwise, %when composing with others,
resulting in more efficient kernels in some cases.

We use templates to separate representation and implementation.
Templates provide a compact tool to iterate over the kernel space and to expose
performance critical tunable parameters. While current template design can not express 
arbitrary optimizations, especially for compute intensive ops, it is enough for 
expressing and tuning memory intensive patterns under study in this work.
\subsection{Code Generation and Kernel Tuning}
\label{section:kernelgen}
\begin{algorithm}[tb]
   \caption{Kernel Generation}
   \label{alg:codegen}
\begin{algorithmic}
   \State {\bfseries Input:} $fusion\_pattern$
   \State {\bfseries Output:} $best\_kernel$
   \State $templates=TemplatesGeneration(fusion\_pattern)$
   \For{$template$ {\bfseries in} $templates$}
     \State {\bfseries Initialize:} $kernel$
     \State $RegisterPlanning(template, fusion\_pattern)$
     \If{$!SharedPlanning(template, fusion\_pattern)$}
       \State $continue$
     \EndIf
     \For{$schedule$ {\bfseries in} $template.schedulelist$}
       \State $clusure=SchedClusure(schedule,fusion\_pattern)$
       \If{$ReductionSchedule(schedule)$}
         \State $ReductionGen(closure,kernel)$
       \ElsIf{$DotSchedule(schedule)$}
         \State $DotGen(closure,kernel)$
       \Else%{$ElemwiseSchedule(schedule)$}
         \State $ElemwiseGen(closure,kernel)$
       \EndIf
     \EndFor % schedule
     \State $KernelEvalUpdate(kernel,best\_kernel)$
   \EndFor % template
\end{algorithmic}
\end{algorithm}
Algorithm \ref{alg:codegen} shows the procedure for kernel generation,
evaluation and performance tuning. It takes as input a computation subgraph 
\emph{fusion\_pattern}, and outputs the optimized \emph{best\_kernel}
after exploring and tuning a series of templates. In practical workloads,
while the number of fusion kernels can be huge, there are only dozens of
unique fusion patterns. It is only necessary to generate a kernel for
each pattern once, and reuse it repetitively.

\emph{TemplatesGeneration} produces a set of templates, each with 
different trade-offs among parallelization, on chip resource requirements
and kernel launch settings. In JIT compilation, it is too time consuming
to tune many templates. We address this issue either by employing a offline
tuning procedure for complex patterns, or by reducing the templates to
a small number with human expert knowledge. This is reasonable for repetitive
DL workloads, since we can optimize once and run many times.

\emph{RegisterPlanning} targets \emph{elementwise} ops whose results are
shared by multiple data dependent ops. We maintain a list of all such ops
in thread local context, avoiding generating code repetitively for them.

\emph{SharedPlanning} optimizes shared memory usage. Either \emph{elementwise},
\emph{reduction}, or \emph{batched-gemm/gemm} ops have reasonable use cases.
However, there are two constraints. The first is the volume constraint: total
shared space allocated should not exceed an upper
threshold $T$. We discuss shared space optimization in 
Section \ref{section:shared-opt}.
%In this work, we set $T$ to be one third of total shared size.
%Please note that shared space can be shared by multiple ops in the same kernel.
%The actual space allocated is usually much less than space requested by 
%each op summed together. We implement a data flow analysis pass based on 
%dominance tree\cite{dominance-tree} to minimize shared space allocation.
The second is layout constraint: we must ensure that for each \emph{op}
with shared allocation, the shared space is only accessed within a single thread
block context. Only when both the volume and layout constraints satisfied can
kernel be generated successfully.

To generate a kernel for a \emph{template}, we traverse its \emph{schedulelist}, 
and generate a code piece for each schedule. Depending on the schedule type
we call separate code emitters. For \emph{elementwise} schedule, a parallel
loop is emitted with thread composition (Figure \ref{fig:composition} (a)),
with special care upon register level results sharing. For \emph{reduction}
schedule, either thread, warp, or block composition is selected depending on 
schedule parameters. For \emph{gemm/batched-gemm} schedule, we also support
either thread or block compositions, depending on the reduction dimension size.
However, we avoid complex tiling loop transformations since our primary focus
here is optimizing memory intensive patterns, not FLOP efficiency.

Each time we generate a kernel. We call \emph{KernelEvalUpdate} to evaluate
its performance and update \emph{best\_kernel} if necessary.
\subsection{Shared Memory Optimization}
\label{section:shared-opt}
\begin{algorithm}[tb]
   \caption{Shared Memory Optimization}
   \label{alg:shared-opt}
\begin{algorithmic}
   \State {\bfseries Input:} $fusion\_pattern$, $req\_map$
   \State {\bfseries Output:} $alloc\_map$
   \State $dom=BuildDominanceTree(fusion\_pattern)$
   \State $topo\_order=TopologicalSort(fusion\_pattern)$
   \For{$inst$ {\bfseries in} $topo\_order$}
     \If{$req\_map.count(inst)$}
       \State {\bfseries Initialize:} $prev\_allocs$
       \For{$operand$ {\bfseries in} $inst.operands()$}
         \State $CollectAllocInfo(operand, prev\_allocs)$
       \EndFor

       \State {\bfseries Initialize:} $shared = False$
       \For{$prev\_inst$ {\bfseries in} $prev\_allocs$}
         \If{$dom.Dominates(inst,prev\_inst)$}
           \If{$!shared$}
             \State $Share(inst, prev\_inst, prev\_allocs)$
             \State $shared = True$
             \State $continue$
           \EndIf
           \State $Reclaim(prev\_inst)$
         \EndIf
       \EndFor

       \If{$!shared$}
         \State $Alloc(inst, alloc\_map)$
       \EndIf
     \Else
       \For{$operand$ {\bfseries in} $inst.operands()$}
         \State $PropagateAllocInfo(inst, operand)$
       \EndFor
     \EndIf
   \EndFor
\end{algorithmic}
\end{algorithm}
Shared memory is the key to compose large granularity of computations effectively
in a kernel. Algorithm \ref{alg:shared-opt} outlines a lightweight shared space 
optimization scheme. Since fusion patterns in our context are typically large,
our goal is not to save space in general, but to constrain worst case shared memory
usage. An example is shown in Figure \ref{fig:example}. Both \emph{dot\_1}
and \emph{add} need $94*94*4$ bytes of shared space. In this graph, the \emph{add} 
can reuse the space allocated for the \emph{dot\_1}. Without optimization,
the hardware occupancy is too low thus compromising performance of the kernel.

A key building block is the dominance tree algorithm\cite{dominance-tree}. 
Instead of using it for control flow analysis, we divert its use here 
in data flow graphs.
The algorithm takes a computation graph and shared memory requests (\emph{req\_map})
as input, and outputs an allocation map (\emph{alloc\_map}). To optimize shared
space sharing, we traverse ops of the computation graph in topological order.
For each op, if no shared space is needed, we simply propagate allocation
information from all its operands. This propagation along data flow edges,
together with the dominance relation of the graph, is vital for shared space 
sharing. Otherwise, if it need shared space for on chip intermediate results transfer,
we merge allocations of all its operands (\emph{CollectAllocInfo}), test the
dominance relation to check
if we can share any pre-allocated space for current op, and reuse the space 
if possible.

\subsection{Implementation}
We implement templates generation, enumeration, and kernel construction 
by replacing the XLA \emph{IREmitter} framework. In particular, we propose and
implement \emph{CUDAEmitter}, a backend to generate CUDA-C code directly
instead of
LLVM IR. We compile CUDA-C with NVCC and integrate the resultant CUBIN
binary with the rest of XLA runtime execution engine to power DL workloads.

\section{Evaluation}
\label{section:evaluation}
%Text of paper \ldots
%\begin{table*}[t]
%\caption{Summary of Evaluation Results}
%\label{tbl:results}
%\begin{center}
%\begin{small}
%%\begin{sc}
%\begin{tabular}{lllllllllll}
%\toprule
%Name & xla/tf-kernel & fs/tf-kernel & xla/tf-perf & fs/tf-perf & elemwise & reduction & gemm & shared-avg & shared-max & alloc/req\\
%\midrule
%nmt & 1.65 & 3.25 & 1.36 & 1.84 & 619 & 296 & 129 & 3.6KB & 18KB & 0.8\\
%multi-interests & 1.63 & 2.27 & 1.25 & 1.57 & 53 & 17 & 21 & 7KB & 36KB & 0.8\\
%rokid & 1.3 & 13.5 & 1.09 & 1.35x & 32 & 18 & 12 & 5.5KB & 16.4KB & 0.9\\
%aiwriter & 2.6 & 5.7 & 2.53 & 4.6 & 3215 & 408 & 1600 & 1.4KB & 4KB & 0.86\\
%logistic & 1.5 & 4.6 & 1.46 & 1.83 & 4 & 3 & 1 & 12B & 12B & 1.\\
%word2vec & 2.9 & 4.9 & 2.6 & 3.3 & 12 & 5 & 3 & 282B & 524B & 0.99\\
%bi-rnn & 2.8 & 6.8 & 2.5 & 4.1 & 151 & 3 & 58 & 0.6 & 3.5 & .98\\
%dyn-rnn & 3.2 & 7.8 & 3.8 & 5.7 & 55 & 4 & 2 & 9B & 16B & 1.\\
%perceptron & 1.1 & 1.3 & 0.94 & 1.18 & 9 & 2 & 5 & 8B & 8B & 1.\\
%var-encoder & 1.5 & 1.9 & 0.8 & 1.09 & 27 & 7 & 9 & 256B & 256B & 1.\\
%\bottomrule
%\end{tabular}
%%\end{sc}
%\end{small}
%\end{center}
%\vskip -0.1in
%\end{table*}
\subsection{Experimental Setup}
We evaluate the fusion optimization and kernel generation mechanisms on four
practical models are six micro-benchmarks. Table \ref{tbl:benchmark-summary}
and Table \ref{tbl:benchmark-characteristics} summarize workload characteristics.
We use the default Tensorflow implementation without compilation,
as well as the XLA compiled version as baselines for our performance 
comparison study. All evaluation results are collected on Nvidia V100 GPUs.
\begin{table}[t]
\caption{Summary of Evaluation Results}
\label{tbl:results}
\begin{center}
\begin{small}
%\begin{sc}
\begin{tabular}{lllllllllll}
\toprule
Name & xla/tf-kernel & fs/tf-kernel & xla/tf-perf & fs/tf-perf\\
\midrule
nmt & 1.65 & 3.25 & 1.36 & 1.84\\
multi-interests & 1.63 & 2.27 & 1.25 & 1.57\\
rokid & 1.3 & 13.5 & 1.09 & 1.35x\\
aiwriter & 2.6 & 5.7 & 2.53 & 4.6\\
logistic & 1.5 & 4.6 & 1.46 & 1.83\\
word2vec & 2.9 & 4.9 & 2.6 & 3.3\\
bi-rnn & 2.8 & 6.8 & 2.5 & 4.1\\
dyn-rnn & 3.2 & 7.8 & 3.8 & 5.7\\
perceptron & 1.1 & 1.3 & 0.94 & 1.18\\
var-encoder & 1.5 & 1.9 & 0.8 & 1.09\\
\bottomrule
\end{tabular}
%\end{sc}
\end{small}
\end{center}
\vskip -0.1in
\end{table}
\subsection{Kernel Launch Savings}
A direct result of computation fusion is reduced number of kernels.
We measure this with the kernel number compression ratio. Table \ref{tbl:results}
shows the results. The column \emph{xla/tf-kernel} shows the kernel number compression
ratio of \emph{xla}, with respect to the kernel number of the Tensorflow baseline. 
Similarly, the \emph{fs/tf-kernel} column shows the kernel number compression ratio 
of our work. Compared to \emph{xla}, the kernel number compression ratio of our 
approach is $1.18$x to $10.38$x higher, with $2.9$x on average.
This is not surprising because we have relaxed fusion conditions by allowing packing
of computations with no data dependences and fusion of \emph{gemm/batched-gemm} with
\emph{elementwise} and \emph{reduction} patterns as well.
\subsection{Performance Speedups}
Larger fusion granularity comes with increased kernel generation complexity. It is
important that aggressive fusion does not compromise GPU efficiency. The last two
columns of Table \ref{tbl:results} show performance results.
Compared to Tensorflow, our approach achieves $1.09$x up to $5.7$x speedups, with 
$2.6$x on average. Compared to \emph{xla}, our approach achieves $1.24$x up to 
$1.84$x speedups, with 1.4x on average.
\subsection{Fusion Patterns Analysis}
\begin{figure}
\centering
\includegraphics[scale=.45]{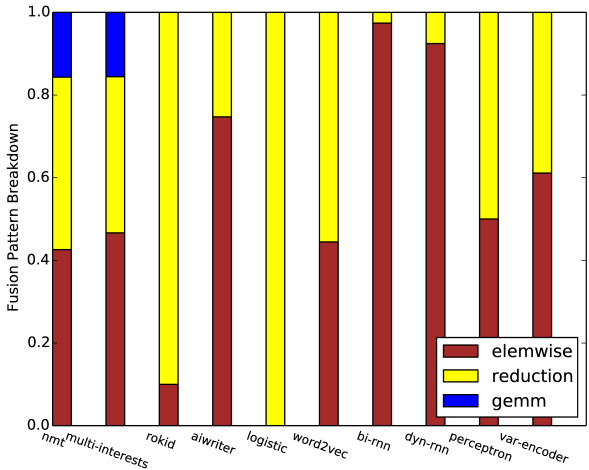}
\caption{Fusion Pattern Composition}
\label{fig:pattern-breakdown}
\end{figure}
We classify a fusion pattern to be one of three categories.
A fusion pattern is \emph{elemwise}, if neither \emph{reduce} nor \emph{gemm} ops
exist in its computation. Otherwise if it contains \emph{gemm} ops, it is marked as
\emph{gemm} pattern. The rest are all marked as \emph{reduction} patterns.
Note if a pattern contains both \emph{gemm} and \emph{reduce}, it is marked as
\emph{gemm} instead of \emph{reduction}.
Figure \ref{fig:pattern-breakdown} shows fusion pattern composition for
all benchmarks. While \emph{elementwise} and \emph{reduction} are common in all
workloads, each workload have different pattern distributions. Both
\emph{nmt} and \emph{multi-interests} have 16\% \emph{gemm} patterns.
%While there are 60\% \emph{elemwise} patterns on average,
%\emph{gemm} and emph{reduction} patterns are also notable, especially in applications
%\emph{nmt}, \emph{multi-interests}, \emph{rokid} and \emph{aiwriter}.

\subsection{Kernel Performance Analysis}
\begin{figure}
\centering
\includegraphics[scale=.45]{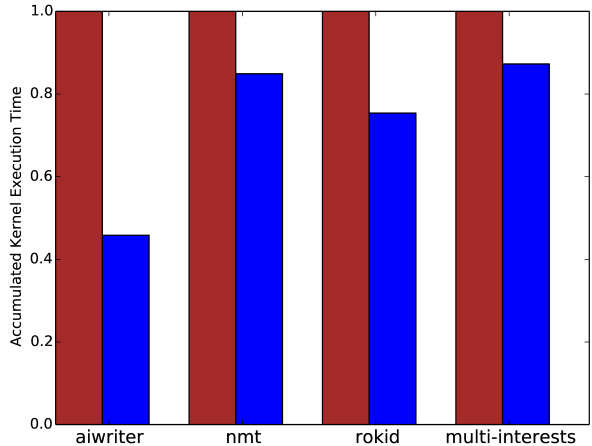}
\caption{Normalized Kernel Performance}
\label{fig:kernel-norm}
\end{figure}

\begin{figure}
\centering
\includegraphics[scale=.45]{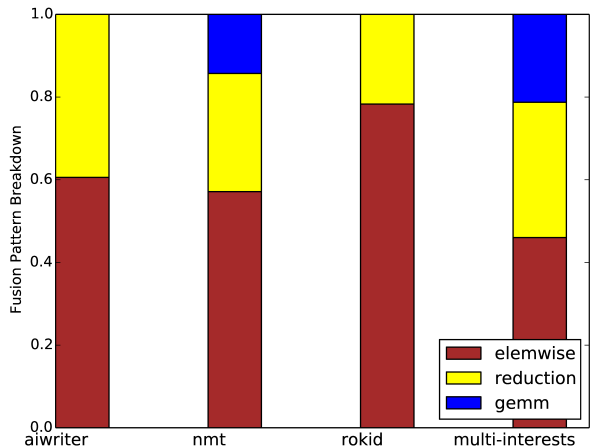}
\caption{Kernel Performance Breakdown}
\label{fig:kernel-breakdown}
\end{figure}
In order to study the kernel performance of our approach. We collect execution time
of all kernels with the \emph{Nvprof} tool for four industry workloads.
These models reflect common requirements of user customized model structures, 
which offer notable potential for large granularity fusion and kernel optimizations.

Figure \ref{fig:kernel-norm} shows the accumulated kernel execution time of all kernels,
normalized to \emph{xla} baseline, for all applications. On average our approach 
achieves 39\% kernel execution time reduction, illustrating the effectiveness of 
fusion and kernel generation towards large granularity.
To further study which fusion patterns contribute most to kernel performance,
Figure \ref{fig:kernel-breakdown} shows the execution time breakdown on three
categories of patterns. Collectively \emph{elemwise} patterns contribute approximately
half of the execution time. The other half is contributed by \emph{reduction} and 
\emph{gemm} patterns together. Please note that \emph{aiwriter} and \emph{rokid}
have \emph{gemm} layers but no \emph{batch-gemm} ops. But the fusion engine does not
fuse them because the granularity is large enough, that calling \emph{cuBLAS} 
routines is more efficient than generating kernels directly. Unlike \emph{aiwriter}
and \emph{rokid}, \emph{nmt} and \emph{multi-interests} have both \emph{gemm} and
\emph{batch-gemm} ops with workload specific shapes which imply interesting 
fusion opportunities. Our fusion engine handles these cases gracefully.
\subsection{Shared Memory Analysis}
\begin{table}[t]
\caption{Shared Memory Statistics}
\label{tbl:shared-memory}
\begin{center}
\begin{small}
%\begin{sc}
\begin{tabular}{llllll}
\toprule
Name & pt-ratio & exe-ratio & shd-avg & max-shd & alloc/req\\
\midrule
nmt & 42\% & 87\% & 3.6KB & 18.6KB & 0.81\\
multiinterests & 24\% & 56\% & 7.5KB & 36KB & 0.88\\
rokid & 90\% & 95\% & 5.5KB & 16.4KB & 0.98\\
aiwriter & 34\% & 62\% & 1.4KB & 4.1KB & 0.87\\
\bottomrule
\end{tabular}
%\end{sc}
\end{small}
\end{center}
\vskip -0.1in
\end{table}
As discussed extensively in Section \ref{section:fusion} and Section \ref{section:codegen},
on chip shared memory is critical when composing numerious computations in a unified kernel.
Table \ref{tbl:shared-memory} shows statistics for four applications.
The \emph{pt-ratio} column shows the percentage of fusion patterns that need shared memory.
The \emph{exe-ratio} column shows the percentage of kernel execution time to which such fusion
patterns contribute. As can be seen, in all applications, shared memory usage is critical 
for fusion and kernel generation.

The shared memory is a precious resource in modern GPUs. How much shared space fusion patterns
actually use? Columns \emph{avg-shd} and \emph{max-shd} show average shared size allocated across 
all patterns, and the maximum size required, respectively. Except \emph{multi-interests}, all
other applications have modest shared memory requirements.

Shared space can be shared among multiple ops in a fused computation. The last column 
\emph{alloc/req} shows the ratio between allocated size and the total space requested. 
The lower the ratio, the higher the shared memory reuse within the kernel. As can be seen, 
the reuse degree is relatively low (high \emph{alloc/req} ratio). 
This seems reasonable, because we not only
fuse \emph{reduction} and \emph{gemm} ops with \emph{elementwise} patterns, but also do packing
of computations with no data dependences. While the averge reuse degree is low, shared space
sharing is critical to reduce worst case shared memory usage. For instance, without sharing,
most demanding fusion patterns of \emph{multi-interests} request up to $72$KB of shared space.
This results in very low execution occupancy, compromising all merits of fusion.
\section{Related Work}
\label{section:relatedwork}
%Text of paper \ldots
GPU kernel fusion, inspired from classical loop optimizations\cite{loopfusion,Kennedy,gao},
is known to boost performance in other application domains.
In database domain, \emph{KernelWeaver}\cite{kernelweaver} proposed
transformations to fuse execution of multiple operators into a single kernel. This work
provided support for both thread and block (CTA) composition of operators, yet with
little support for tuning of implementation schedules. In the HPC domain, \cite{hpcfusion}
formulated GPU kernel fusion as an combinatorial search problem, and searched the
solution space for an optimized fused kernel. In image processing domain,
\cite{cgo2019,scopes} formulated the image pipeline fusion as a graph cut problem. 
For machine learning workloads, \cite{ppopp2015} proposed a kernel fusion technique to
generate efficient kernels for a specific computation pattern. 
The \emph{XLA} compilation framework\cite{xla} can handle more general computation pattern, 
but offers only basic capability for fusion and kernel generation. 
However, XLA relies on empirical rules to encode fusion opportunities, and does not support
fusion and code generation of composition of \emph{elementwise}, \emph{reduction}
and \emph{gemm} patterns.

The separation of optimization specification and implementation is important for
performance modeling and tuning. There are extensive works for compute intensive
operators, such as Halide\cite{halide}, TVM\cite{tvm,tvmgit}, and TensorComprehensions\cite{tc}.
The implementation template in our work targets large, complex memory intensive 
computation patterns.

There are recent advances on code generation of compute intensive DNN layers.
\cite{pbqp} proposed a solution for selecting fast kernel
implementations in the global context by formulating it as a PBQP
problem. Boda\cite{boda} is a code generator that generates code for CNN layers on mobile
platforms. Latte\cite{latte} is a DSL system for DNN allowing users to specify, synthesize
and optimize code for NN layers. SLINGEN\cite{slingen} is another DSL system which takes
mathematical specifications and generates optimized C functions for linear algebra
operators with small input sizes.
These research are relevant but complementary to our work.
\section{Conclusion}
\label{section:conclusion}
%Text of paper \ldots
Fine grained memory intensive computations are abundant in DL workloads. 
This work tackles this problem from two aspects. First, we propose a novel computation
fusion framework to explore and optimize fusion plans. In particular, we propose an ILP
formulation for fusion plans optimizations. Our fusion framework not only supports
composition of \emph{elementwise} and \emph{reduction} ops with or without data dependences,
but also support composition of \emph{gemm/batched-gemm} ops, thus enabling collective
optimizations of both compute and memory intensive computations.

Together with fusion plan optimizations, we propose a code generation algorithm to produce
optimized kernels for GPUs. With extensive intermediate results sharing via either 
registers or shared memory, our work is capable of supporting very large fusion granularity.
Experimental results on six benchmarks and four industry scale practical models are
encouraging. Overall, \emph{FusionStitching} can reach up to 5.7x speedup compared
to Tensorflow baseline, and achieves
1.25x to 1.85x performance speedups compared to current state of the art, with 1.4x on
average (geometric mean).
%% Acknowledgments
%\begin{acks}                            %% acks environment is optional
%                                        %% contents suppressed with 'anonymous'
%  %% Commands \grantsponsor{<sponsorID>}{<name>}{<url>} and
%  %% \grantnum[<url>]{<sponsorID>}{<number>} should be used to
%  %% acknowledge financial support and will be used by metadata
%  %% extraction tools.
%  This material is based upon work supported by the
%  \grantsponsor{GS100000001}{National Science
%    Foundation}{http://dx.doi.org/10.13039/100000001} under Grant
%  No.~\grantnum{GS100000001}{nnnnnnn} and Grant
%  No.~\grantnum{GS100000001}{mmmmmmm}.  Any opinions, findings, and
%  conclusions or recommendations expressed in this material are those
%  of the author and do not necessarily reflect the views of the
%  National Science Foundation.
%\end{acks}

%% Bibliography
\bibliography{paper.bib}

%% Appendix
%\appendix
%\section{Appendix}

%Text of appendix \ldots

\end{document}